\documentclass[runningheads]{svmult}
\usepackage{makeidx}   
\usepackage{graphicx}  
\usepackage{subeqnar}  
\usepackage{multicol}  
\usepackage{physprbb}  
\makeindex             

\usepackage{amssymb}   
\newcommand{\Bx}{x_{\rm B}}

\newcommand{\cQ}{{\cal Q}}
\newcommand{\re}{\Re\mbox{e}}
\newcommand{\im}{\Im\mbox{m}}
\newcommand{\GeV}{\mbox{GeV}}

\begin{document}
\title*{Predictions for deeply virtual Compton scattering on a spin-one
target}
\toctitle{Predictions for deeply virtual Compton scattering on a spin-one
target}
%
%
\titlerunning{Predictions for DVCS}
%
\author{Axel Kirchner\inst{1} \and Dieter M\"uller\inst{2}
}
\authorrunning{Axel  Kirchner \and  Dieter M\"uller }
%
%
\institute{
Institut f\"ur Theoretische Physik, Universit\"at Regensburg,
D-93040 Regensburg, Germany
\and
Fachbereich Physik, Universit\"at  Wuppertal, 
D-42097 Wuppertal, Germany
}

\maketitle              

\begin{abstract}
We consider hard leptoproduction of a photon on a spin-one
target and give the general azimuthal angular dependence of the differential
cross section. Furthermore, we estimate the beam spin asymmetry for an unpolarized
deuteron target at HERMES.
\end{abstract}

\section{Introduction}

Exclusive two-photon processes in the light-cone dominated region, i.e., in
the generalized Bjorken limit, are most suitable for the exploration of the
partonic content in hadrons, since in leading order (LO) both photons directly
couple to one quark line \cite{MueRobGeyDitHor94}. In such processes one can
measure, for instance, different photon-to-meson form factors, i.e.,
$\gamma^\ast \gamma \to M$, the production of hadron pairs and also
processes like $\gamma^\ast N \to N \gamma $ or $\gamma N \to N l^+ l^- $.
The latter two are denoted as deeply virtual Compton scattering (DVCS) in
the space- and time-like regions
\cite{Rad96Ji96a,DieGouPirRal97,BelMueNieSch00,BelMueKirSch00,BerDiePir01,BelMueKir01}.

The factorization of short- and long-range dynamics is formally given by the
operator product expansion (OPE) of the time ordered product of two
electromagnetic currents, which has been worked out at leading twist-two in
next-to-leading order (NLO) and at twist-three level in LO  of
perturbation theory (for references see \cite{BelMueKir01}).
However, one should be aware that the partonic
hard-scattering part, i.e., the Wilson coefficients, contains collinear
singularities, which are absorbed in the non-perturbative distributions by a
factorization procedure, which has been proven at twist-two level
\cite{ColFre98JiOsb98}.

The non-perturbative distributions are defined in terms of
light-ray operators with definite twist sandwiched between the corresponding
hadronic states. These process dependent correlation functions are sensitive to
different aspects of hadronic physics. Especially, in DVCS one can access
the so-called generalized parton distributions (GPDs). The second moment of
the flavour singlet GPDs is related to the expectation value of the
energy momentum tensor. Thus, it gives in principle information on the angular
orbital momentum fraction of the nucleon spin carried by quarks \cite{Ji96}.
We should stress that this process is a new tool to probe
the partonic content of the nucleon on the level of amplitudes and, thus,
it provides us new information \cite{RalPir01}.

Recently, the DVCS process has been measured by the H1 collaboration
\cite{Adletal01} in the small $\Bx$ region (see also \cite{Sau00}) as well
as in single beam spin asymmetries by the HERMES \cite{Airetal01} and CLAS
\cite{Steetal01} collaborations. The theoretical predictions depend on the
GPDs, which are unknown. However, they are constrained by sum rules and the
reduction to the parton densities in the forward kinematics. All
experimental data are consistent with an oversimplified model, which
satisfies the constraints and, thus, also with each other
\cite{BelMueKir01}.

It is appealing to employ DVCS for the investigation of other hadrons and
nuclei. An appropriate candidate is deuteron, which has been widely used as
a target in lepto-scattering experiments. This nucleus has been extensively
studied in both deep-inelastic \cite{HooJafMan89} and elastic
\cite{BroJiLep83,GilGro01GarOrd01} scattering. From the theoretical point
of view it would be desired to have complementary information, which could
give us a deeper understanding of this nucleus in terms of its fundamental
degree of freedom \cite{BerCanDiePir01}.

In this paper we outline the OPE approach applied to DVCS. Moreover, we
determine the azimuthal angular dependence of the cross section for a spin-1
target. Relying on qualitative properties of
GPDs, which are consistent with the DVCS data for the proton target, we
estimate the size of the beam spin asymmetry for HERMES kinematics.

\section{General formalism}

For the leptoproduction of a photon
\begin{eqnarray}
l^\pm (k) h (P_1) \to l^\pm  (k^\prime) h (P_2) \gamma (q_2)
\end{eqnarray}
on a hadronic target $h$ with the mass $M$ the five-fold cross section 
\begin{eqnarray}
\label{WQ}
\frac{d\sigma}{d\Bx dy d|\Delta^2| d\phi d\varphi}
=
\frac{\alpha^3  \Bx y } {16 \, \pi^2 \,  {\cal Q}^2 \sqrt{1 + \epsilon^2}}
\left| \frac{\cal T}{e^3} \right|^2 , \quad
\epsilon \equiv 2 \Bx \frac{M}{{\cal Q}},
\end{eqnarray}
depends on the Bjorken variable $\Bx = {\cal Q}^2 / 2 P_1\cdot q_1 $,
where ${\cal Q}^2 = -q_1^2$ with $q_1 = k - k'$, the momentum transfer
square $\Delta^2 = (P_2 - P_1)^2$, the lepton energy
fraction
$y= P_1\cdot q_1/P_1\cdot k$ and, in general, two
azimuthal angle. 
We use the target rest frame,
where the virtual photon three-momentum 
points towards the negative $z$-direction. 
$\phi$
is the angle between the lepton and hadron scattering planes and $\varphi =
{\mit\Phi} - \phi_N$ is the difference of the azimuthal angle $\mit\Phi$ of
the spin vector \begin{eqnarray} \label{Def-SpiVec} S^\mu = (0,
\cos{\mit\Phi} \sin{\mit\Theta}, \sin{\mit\Phi} \sin{\mit\Theta},
\cos{\mit\Theta}) \end{eqnarray} and the azimuthal angle $\phi_N$ of the
recoiled hadron (see Ref.\ \cite{BelMueKir01}).

In the following we consider this process in the  (generalized) Bjorken limit,
$\cQ^2 \sim P_1 \cdot q_1 =$ large, $\Delta^2$ and $M^2$ are comparably small. 
The amplitude ${\cal T}$ is the sum of the DVCS ${\cal T}_{\rm DVCS}$ and
Bethe-Heitler (BH) ${\cal T}_{\rm BH}$ amplitudes:
\begin{equation}
\label{Def-T2}
{\cal T}^2
= \sum_{\lambda^\prime, \Lambda^\prime} \left\{
|{\cal T}_{\rm BH}|^2 + |{\cal T}_{\rm DVCS}|^2 + {\cal I} \right\},
\end{equation}
with the interference term
\begin{equation}
{\cal I}
= 
{\cal T}_{\rm DVCS} {\cal T}_{\rm BH}^\ast
+ {\cal T}_{\rm DVCS}^\ast {\cal T}_{\rm BH},
\end{equation}
where the recoiled lepton $(\lambda^\prime)$ and
hadron $(\Lambda^\prime)$ polarization will not be observed.

Each of these three terms in Eq.\ (\ref{Def-T2}) can be 
calculated in a straightforward manner by the
contraction of a known leptonic tensor $L_{\mu\dots}$ with the
DVCS tensor $T^{\rm DVCS}_{\mu\nu}$  or/and hadronic current
$J_\alpha$. The former one is given in terms of GPDs, while the latter one
is parametrized in terms of elastic electromagnetic form factors.
For instance, the interference term reads
\begin{eqnarray}
\label{Con-LepHad}
\sum_{\lambda^\prime}
{\cal I} = \frac{\pm e^6}{{\cal Q}^2 \Delta^2} L^{\mu\nu\alpha}(k,k^\prime)
T^{\rm DVCS}_{\mu \nu}(P,\Delta,q)  J^\dagger_\alpha(\Delta)  + \mbox{h.c.}
\quad \left\{ {+ \mbox{\ for\ } e^- \atop - \mbox{\ for\ } e^+} \right.  ,
\end{eqnarray}
where $P = P_{1} +P_{2}$ and $q=(q_1+q_2)/2$. The resulting
predictions for the spin-0 and -1/2 targets are presented at the twist-three level
in Refs.\ \cite{BelMueKirSch00,BelMueKir01}.

Let us now consider the spin-1 target in more detail.
The hadronic current
\begin{eqnarray}
\label{Def-EM-Cur-1}
J_\mu = - \epsilon_2^\ast\! \cdot\! \epsilon_1 P_\mu\, G_1
+ \left(\epsilon_2^\ast\! \cdot\! P \epsilon_{1\mu} +
 \epsilon_1\! \cdot\! P \epsilon_{2\mu}^\ast \right) G_2  -
\epsilon_2^\ast\! \cdot P\,  \epsilon_1\! \cdot\! P 
\frac{P_\mu}{2 M^2}\,  G_3
\end{eqnarray}
is given by three form factors $G_i(\Delta^2)$ with $i=\{1,2,3\}$, 
where $\epsilon_{1\mu}$
($\epsilon_{2\mu}$) denote the three polarization vectors for the initial
(final) hadron. The form factors $G_i(\Delta^2)$ have been measured and 
their parametrizations are available in the literature,
see Ref.\ \cite{GilGro01GarOrd01} and references therein.

The DVCS hadronic tensor is given by the time-ordered product of the
electromagnetic currents $j_\mu$,
which is sandwiched
between hadronic states with different momenta. In LO of
perturbation theory and at twist-two accuracy it reads \cite{BelMueNieSch00}
\begin{eqnarray}
\label{HadronicTensor}
T_{\mu\nu} (\xi, \Delta^2, {\cal Q}^2) &=&
\frac{i}{e^2} \int dx {\rm e}^{i x \cdot q}
\langle P_2 | T j_\mu (x/2) j_\nu (-x/2) | P_1 \rangle \\
&=&
- {\cal P}_{\mu\sigma} g_{\sigma\tau} {\cal P}_{\tau\nu}
\frac{q \cdot V_1}{P \cdot q}
- {\cal P}_{\mu\sigma} i\epsilon_{\sigma \tau q \rho} {\cal P}_{\tau\nu}
\frac{A_{1\, \rho}}{P \cdot q}
, \nonumber
\end{eqnarray}
where the projection operator
$
{\cal P}_{\mu\nu} = g_{\mu\nu} - q_{1 \mu} q_{2 \nu}/{q_1 \cdot q_2}
$
ensures gauge invariance. For convenience we introduced the scaling variable
$\xi\approx  \frac{\Bx}{2 - \Bx}$.
At twist-two level the amplitudes $V_1$ and $A_1$ can be decomposed in a
complete basis of nine Compton form factors (CFFs). Adopting the notation of
Ref.\ \cite{BerCanDiePir01}, they read in the vector case
\begin{eqnarray}
\label{Def-V1-Sp1}
V_\mu &=& - \epsilon_2^\ast\! \cdot\! \epsilon_1 P_\mu\, {\cal H}_1
+ \left(\epsilon_2^\ast\! \cdot\! P \epsilon_{1\mu} +
 \epsilon_1\! \cdot\! P \epsilon_{2\mu}^\ast \right) {\cal H}_2
-\epsilon_2^\ast\! \cdot P\,  \epsilon_1\! \cdot\! P 
\frac{P_\mu}{2 M^2}\,  {\cal H}_3 \\
&&\!\!\!
+ \left(\epsilon_2^\ast\! \cdot\! P \epsilon_{1\mu} -
 \epsilon_1\! \cdot\! P \epsilon_{2\mu}^\ast \right) {\cal H}_4
+
\left(\frac{2 M^2 \left\{
\epsilon_2^\ast\! \cdot\! q \epsilon_{1\mu} +
 \epsilon_1\! \cdot\! q \epsilon_{2\mu}^\ast \right\}}{P\cdot q}
 + \frac{\epsilon_2^\ast\! \cdot\! \epsilon_{1}}{3}
 P_\mu \right) {\cal H}_5,
\nonumber
\end{eqnarray}
and in the axial-vector case
\begin{eqnarray}
\label{Def-A1-Sp1}
A_\mu &=&
i \epsilon_{\mu \epsilon_2^\ast  \epsilon_1 P } \widetilde {\cal H}_1
- \frac{i\epsilon_{\mu \Delta P \epsilon_1}\, \epsilon_2^\ast\! \cdot\! P+
i \epsilon_{\mu \Delta P \epsilon_2^\ast}\, \epsilon_1\! \cdot\! P}{M^2}
 \widetilde {\cal H}_2
\\
 & &\!\!\!
- \frac{i\epsilon_{\mu \Delta P \epsilon_1}\, \epsilon_2^\ast\! \cdot\! P-
i \epsilon_{\mu \Delta P \epsilon_2^\ast}\, \epsilon_1\! \cdot\! P}{M^2}
 \widetilde {\cal H}_3
- \frac{i\epsilon_{\mu \Delta P \epsilon_1}\, \epsilon_2^\ast\! \cdot\! q+
i \epsilon_{\mu \Delta P \epsilon_2^\ast}\, \epsilon_1\! \cdot\! q}{q\! \cdot\! P}
\widetilde {\cal H}_4,
\nonumber
\end{eqnarray}
where $1/\cQ$-power suppressed effects have been neglected. The remaining
logarithmical $\cQ$-dependence is governed by perturbation theory.

The CFFs in Eqs. (\ref{Def-V1-Sp1}) and (\ref{Def-A1-Sp1}) are
given by a convolution of perturbatively calculable coefficient functions
$C^{(\pm)}$ and twist-two GPDs via
\begin{eqnarray}
\label{DefTw2}
&&
{\cal H}_k (\xi)
= \sum_{i=u,\dots}
\int_{- 1}^{1} \! dx \, C_i^{(-)} (\xi, x) H_k^i(x, \eta)_{|\eta=-\xi},
\quad \mbox{for}\quad k=\{1,\dots,5\},
 \\
&&
\widetilde {\cal H}_k (\xi)
= \sum_{i=u,\dots}\int_{- 1}^{1} \! dx \, C_i^{(+)} (\xi, x)
\widetilde H^i_k(x, \eta)_{|\eta=-\xi},
\quad \mbox{for}\quad k=\{1,\dots,4\} .
\end{eqnarray}
For each quark species $i$ we have 
nine GPDs. The two sets $\{H^i_1,\dots,H^i_5\}$ and $\{\widetilde{H}^i_1,\dots,
\widetilde{H}_4^i\}$
are defined by off-forward matrix elements of vector and axial-vector 
light-ray operators.  The coefficient functions $C^{(\mp)}$ have perturbative
expansion.
In LO they read for the even ($-$) and odd ($+$) parity sectors
\begin{equation}
\label{CoeffFunction}
\xi\, {C_{(0)i}^{(\mp)}} \left( \xi, x \right)
= \frac{Q^2_i}{1 - x/\xi - i 0}
\mp
\frac{Q^2_i}{1 + x/\xi - i 0},
\end{equation}
where $Q_i$ is the fractional quark charge.

Employing the parametrizations (\ref{Def-EM-Cur-1}), (\ref{Def-V1-Sp1}) and
(\ref{Def-A1-Sp1}), the contractions of leptonic and hadronic tensors 
provide the kinematically exact expression for the squared BH term
(of course, in tree approximation), while the interference term
\begin{eqnarray}
\label{Def-T2-Int}
&&\hspace{-0.5cm}\sum_{\lambda^\prime}
{\cal T}_{\rm BH} {\cal T}_{\rm DVCS}^\ast
=
\\
&& \frac{2 - 2y + y^2}{y^2 {\cal P}_1(\phi){\cal P}_2(\phi) }
 \frac{ 4  \xi}{\Delta^2 {\cal Q}^4}
 \left( k^\sigma -\frac{q^\sigma}{y}  \right)
\left[
\left( J_\sigma + 2 \Delta_\sigma \frac{q\!\cdot\!J}{{\cal Q}^2} \right)
q\!\cdot\!V_1^\dagger
+ 2 i \epsilon_{\sigma q \Delta J} \frac{q\!\cdot\!A_1^\dagger}{{\cal Q}^2} \right]
\nonumber\\
&&\!\!\!\! +\frac{\lambda(2 - y) y}{y^2 {\cal P}_1(\phi){\cal P}_2(\phi)}
 \frac{4 \xi}{\Delta^2 {\cal Q}^4}
\left( k^\sigma -\frac{q^\sigma}{y}  \right)
\left[ 
\left( J_\sigma + 2 \Delta_\sigma \frac{q\!\cdot\!J}{{\cal Q}^2} \right)
q\!\cdot\! A_1^\dagger
+ 2 i \epsilon_{\sigma q \Delta J} \frac{q\!\cdot\!V_1^\dagger}{{\cal Q}^2}
\right],
\nonumber
\end{eqnarray}
and the squared DVCS amplitude
\begin{eqnarray}
\label{Def-T2-DVCS}
\sum_{\lambda^\prime} |{\cal T}_{\rm DVCS}|^2 &=&
8 \frac{2 - 2y + y^2}{y^2} \frac{\xi^2}{{\cal Q}^6}
\left(q\!\cdot\! V_1\ q\!\cdot\! V_1^\dagger +
q \!\cdot\! A_1\ q \!\cdot\! A_1^\dagger \right)
\nonumber \\
&&+ 8 \frac{\lambda (2-y)}{y} \frac{\xi^2}{{\cal Q}^6}
\left(q\!\cdot\! V_1\ q\!\cdot\! A_1^\dagger +
q\!\cdot\! A_1\ q\!\cdot\! V_1^\dagger  \right)
\end{eqnarray}
are expanded with respect to $1/\cQ$.
In contrast to the squared DVCS amplitude the interference as well as the
squared BH terms have an additional $\phi$-dependence due to the (scaled) BH
propagators
\begin{equation}
\label{Par-BH-Pro}
{\cal P}_1
\approx - \frac{1}{y} \left\{ 1-y + 2 K \cos(\phi) \right\}
\, , \qquad
{\cal P}_2
\approx
\frac{1}{y}
\left\{1  + 2 K \cos(\phi)
\right\}
.
\end{equation}
The kinematical factor
\begin{equation}
K \approx \sqrt{ -\frac{\Delta^2}{{\cal Q}^2} (1 - \Bx) ( 1 - y )
\left( 1 - \frac{\Delta^2_{\rm min}}{\Delta^2} \right)}.
\end{equation}
vanishes at the kinematical boundary $\Delta^2 = \Delta_{\rm min}^2
\approx M^2 \Bx^2/(1 - \Bx)$.

\section{Observables for leptoproduction of a photon}

In our frame the contractions of leptonic and hadronic tensors, see Eq.\
(\ref{Con-LepHad}), yield finite sums of Fourier harmonics, whose form is
governed by general principles. After summation over the final
polarization states, which is not indicated below, the three parts of the
squared amplitude read for massless leptons \cite{BelMueKir01}:
\begin{eqnarray}
\label{Par-BH}
&&\hspace{-0.8cm}|{\cal T}_{\rm BH}|^2
= \frac{e^6 (1 + \epsilon^2)^{-2}}
{\Bx^2 y^2  \Delta^2\, {\cal P}_1 (\phi) {\cal P}_2 (\phi)}
\left\{
c^{\rm BH}_0
+  \sum_{n = 1}^2
\left[ c^{\rm BH}_n \, \cos{(n\phi)}
 + s^{\rm BH}_n \, \sin{(\phi)}\right]
\right\} \, ,
\\
\label{AmplitudesSquared}
&&\hspace{-0.8cm} |{\cal T}_{\rm DVCS}|^2
=
\frac{e^6}{y^2 {\cal Q}^2}\left\{
c^{\rm DVCS}_0
+ \sum_{n=1}^2
\left[
c^{\rm DVCS}_n \cos (n\phi) + s^{\rm DVCS}_n \sin (n \phi)
\right]
\right\} \, ,
\\
\label{InterferenceTerm}
&&\hspace{-0.8cm}{\cal I}
= \frac{\pm e^6}{\Bx y^3 {\cal P}_1 (\phi) {\cal P}_2 (\phi) \Delta^2}
\left\{
c_0^{\cal I}
+ \sum_{n = 1}^3
\left[
c_n^{\cal I} \cos(n \phi) +  s_n^{\cal I} \sin(n \phi)
\right]
\right\} \, ,
\end{eqnarray}
where the $+$ ($-$) sign in the interference stands for the negatively
(positively) charged lepton beam.

The Fourier coefficients can be calculated from Eqs.\
(\ref{Def-T2-Int}), (\ref{Def-T2-DVCS}), and an analogous one for the squared BH
amplitude by summing over the polarization $\Lambda^\prime$, where we
can employ the common projector technique. For a spin-1 particle we
have (see for instance \cite{Pas83})
\begin{eqnarray}
\label{Def-PolSum}
\epsilon_{1\mu}^\ast(\Lambda=0) \epsilon_{1\nu} (\Lambda=0) &=& S_\mu S_\nu,
\\
\epsilon_{1\mu}^\ast(\Lambda=\pm 1) \epsilon_{1\nu} (\Lambda=\pm 1) &=&
\frac{1}{2} \left(-g_{\mu\nu} + \frac{P_{1\mu}P_{1\nu}}{M^2} 
- S_\mu S_\nu + \frac{i \Lambda}{M} \epsilon_{\mu\nu P_1 S}  \right),
\nonumber
\end{eqnarray}
where $\Lambda=\{+1,0,-1\}$ denotes the magnetic quantum number with respect
to the quantization direction given by the spin vector $S_\mu$
defined in Eq.\ (\ref{Def-SpiVec}).
Obviously, the spin sum of the recoiled hadron  is
\begin{eqnarray}
\sum_{\Lambda^\prime=-1}^1
\epsilon_{2\mu}^\ast(\Lambda^\prime) \epsilon_{2\nu} (\Lambda^\prime) =
-g_{\mu\nu} + \frac{P_{2\mu}P_{2\nu}}{M^2} .
\end{eqnarray}

As we see, for a spin-1 target the Fourier coefficients quadratically  depend 
on the spin vector $S_\mu$ and, thus, we find the following decomposition
\begin{eqnarray}
\label{Def-FouDec}
c_n^{T} &=& c_{n,{\rm unp}}^{T} +
c_{n,{\rm LP}}^{T} \cos({\mit\Theta}) +
c_{n,{\rm TP}}^{T}(\varphi) \sin({\mit\Theta})
+
c_{n,{\rm LTP}}^{T}(\varphi) \sin(2{\mit\Theta})
\nonumber\\
& & +
c_{n,{\rm LLP}}^{T} \cos^2({\mit\Theta})  +
c_{n,{\rm TTP}}^{T}(\varphi) \sin^2 ({\mit\Theta})
\end{eqnarray}
for $ T=\{{\rm BH}, {\cal I}, {\rm DVCS}\}$.
An analogous decomposition holds true for the odd harmonics $s_n^{T}$.
The unpolarized and the longitudinally polarized coefficients
$c/s_{n,{\rm unp}}^{T}$, $c/s_{n,{\rm LP}}^{T}$,  and $c/s_{n,{\rm LLP}}^{T}$,
respectively, are independent of $\varphi$. The  transverse
coefficients 
$c/s_{n,{\rm TP}}^{T}$  and 
the transverse-longitudinal interference terms 
 can be decomposed with
respect  to the first harmonics in the azimuthal angle $\varphi$
\begin{eqnarray}
c_{n,{\rm TP}}^{T}(\varphi)  &=& c_{n,{\rm TP}+}^{T} \cos(\varphi) +
s_{n,{\rm TP}-}^{T} \sin(\varphi),\\
c_{n,{\rm LTP}}^{T}(\varphi) &=& c_{n,{\rm LTP}+}^{T} \cos(\varphi) +
s_{n,{\rm LTP}-}^{T} \sin(\varphi),
\end{eqnarray}
while $c/s_{n,{\rm TTP}}^{T}$ may be written as
\begin{eqnarray}
c_{n,{\rm TTP}}^{T}(\varphi)  &=&
c_{n,{\rm TTP}\Sigma}^{T}+
 c_{n,{\rm TTP}\Delta}^{T} \cos(2\varphi) +
s_{n,{\rm TTP}\pm}^{T} \sin(2\varphi).
\end{eqnarray}
Analogous equations hold true for the odd harmonics (just replace 
$c \leftrightarrow s$). Let us add that with this notation the 
$c(s)$ harmonics are given by the real (imaginary) part of certain
linear combinations of form factors and/or CFFs.

Altogether we have for a given harmonic in $\phi$ nine possible
observables\footnote{Note that $c_{n,{\rm TTP}\Sigma}^{T}$ does not belong
to an independent frequency, rather it can be included
in the constant and  $\cos^2(\theta)$ terms of Eq.\ (\ref{Def-FouDec}). }.
 In principle, they can be measured by an appropriate adjustment
of the spin vector and Fourier analysis with respect to the azimuthal angle
$\varphi$. The interference term linearly depends on the CFFs
and is, thus, of special interest. In facilities that have both kinds of
leptons it can be separated by means of the charge asymmetry. The dominant
harmonics are $c/s_1^{\cal I}$, predicted at leading twist-two. We write
them as product of leptonic prefactors $\cal L$ and `universal' functions
${\cal C}_{i}^{\cal I}$:
\begin{eqnarray}
\left\{ c_{1,i}^{\cal I} \atop s_{1,i}^{\cal I}   \right\} =
 \left\{{\cal L}_{1,i}^{{\cal I}c} \atop {\cal L}_{1,i}^{{\cal I}s} \right\}
\left\{\re  \atop \im \right\}  {\cal C}_{i}^{\cal I},
\quad \mbox{for}\
i=\{{\rm unp},  \cdots,{\rm TTP}-\}.
\end{eqnarray}
As mentioned before, they depend on nine CFFs
\begin{eqnarray}
{\cal C}_i^{\cal I} =
\left( G_1 \cdots G_3 \right) {\cal M}_i^{\cal I}
\left(
\begin{array}{c}
{\cal H}_1\\ \vdots \\ \widetilde{{\cal H}}_4
\end{array}
\right),
\end{eqnarray}
where the matrix ${\cal M}$ will be presented elsewhere \cite{KirMue02a}.
Single spin-flip and unpolarized as well as double spin-flip contributions provide
the imaginary and real part, respectively, of the nine linear combinations
${\cal C}^{\cal I}_i$.

\section{Estimate of the beam-spin asymmetry}

In this section we estimate the size of the beam spin asymmetry
\begin{eqnarray}
\label{Def-ALU}
A_{\rm LU}(\phi)
=
\frac{
d\sigma^\uparrow(\phi) - d\sigma^\downarrow(\phi)
}{
d\sigma^\uparrow (\phi)+ d\sigma^\downarrow(\phi)
}\, ,
\end{eqnarray}
at large $y$ and small momentum transfer for the HERMES experiment.
We should also ensure that $M^2/\cQ^2$ corrections are under control. Since
$M= 1875.6$ MeV we  require $\cQ^2 \ge 4\  \GeV^2$.
Because of
large $y$, the BH amplitude dominates the DVCS one and thus the beam spin
asymmetry is approximately determined by
\begin{eqnarray}
A_{\rm LU}(\phi)
\sim
\pm \frac{\Bx}{y}
\frac{s_{1,{\rm unp}}^{\cal I}}{c_{0,{\rm unp}}^{\rm BH}} \sin(\phi)
\quad\mbox{with}
\quad \left\{ {+ \mbox{\ for\ } e^- \atop - \mbox{\ for\ } e^+} \right. .
\end{eqnarray}
For $-\Delta_{\rm min} \ll -\Delta^2 \ll M^2$ and $\Bx \lesssim 0.3 $ the Fourier coefficients can be
drastically simplified due a crude approximation of kinematical factors
\begin{eqnarray}
\label{App-ALU}
A_{\rm LU}(\phi) \sim \pm \sqrt{\frac{-\Delta^2}{\cQ^2}(1-y)}\,
\frac{\Bx\im {\cal H}_1(\xi,\Delta^2,\cQ^2)_{|\xi=\Bx/2}}{G_1(\Delta^2)}
\sin(\phi) .
\end{eqnarray}
For a spin-1/2 target one finds the analogous equation, i.e., 
$G_1 \to F_1$ and ${\cal H}_1 \to {\cal H}$. Note that the sume rule
\begin{eqnarray}
\int_{-1}^{1}dx \left(Q_u
\left\{\!\!{H_1 \atop H}\!\!\right\}_{u_v}\!\!\!\!\!\!\! (x,\xi,\Delta^2,\cQ^2)+
Q_d \left\{\!\!{H_1 \atop H}\!\!\right\}_{d_v}\!\!\!\!\!\!\! (x,\xi,\Delta^2,\cQ^2)
\right)=
\left\{\!\! {G_1 \atop F_1}\!\!\right\}(\Delta^2)
\end{eqnarray}
suggest that the $\Delta^2-$dependence of the
valence-quark GPDs is essentialy given by
$G_1$ and $F_1$, respectively. The analyses of the H1, HERMES and CLAS
data at LO indicate that the $\Delta^2$ fall-off of the sea-quark GPDs
is steeper than the valence quark ones. Thus, we
neglect the sea quark contribution and write the valence like
GPDs as a product of form factor and quark distribution function
\begin{eqnarray}
\label{App-CFFH1}
\frac{\im {\cal H}(\xi,\Delta^2)}{F_1(\Delta^2)} &\sim &
  \pi \left\{ Q_u^2  q_{u_v}(\xi)+Q_d^2  q_{d_v}(\xi) \right\}_{|\xi=\Bx/2},
\\
\frac{\im {\cal H}_1(\xi,\Delta^2)}{G_1(\Delta^2)} &\sim &
  \pi
\frac{Q_u^2+Q_d^2}{2}\left\{q_{u_v}(\xi)+q_{d_v}(\xi)\right\}_{|\xi=\Bx/2}.
\end{eqnarray}
For HERMES kinematics with central values $\langle \Bx\rangle =0.11$,
$\langle \Delta^2\rangle =-0.27\ \GeV^2$ and $\langle \cQ^2\rangle=2.6\
\GeV^2$ we find with the MRS A$^\prime$ parametrization $A_{\rm
LU}(\phi)\sim 0.3 \sin(\phi)$ for positron-proton scattering, which is
consistent with experimental data \cite{Airetal01}.
For $\cQ^2 =4\ \GeV^2$, $\Bx =0.1$ and $ \Delta^2 =-0.3\
\GeV^2$ we estimate
\begin{eqnarray}
A_{\rm LU}(\phi)&\sim & - 0.13 \sin(\phi) \qquad \mbox{for deuteron target},
\\
A_{\rm LU}(\phi)&\sim & - 0.16 \sin(\phi) \qquad \mbox{for proton target}.
\nonumber
\end{eqnarray}

\section{Summary}

We gave the general azimuthal angular dependence of the leptoproduction
cross section of a photon on a spin-1 target. Such experiments allow to
study the deuteron from a new perspective. Compared to DIS or elastic
lepton-deuteron scattering, they provide additional information, contained
in the CFFs. 

At twist-two level there are nine CFFs. In the case of a polarized beam and
target, with an adjustable quantization direction, the imaginary and real
part of all these CFFs can be separately measured by means of the charge
asymmetry. Moreover, an appropriate Fourier analysis allows to eliminate the
twist-three contamination. In this way one gets the maximal access to
the deuteron GPDs at twist-two level, with a contamination of $M^2/\cQ^2$
and $\Delta^2/\cQ^2$ contributions.

As discussed, for certain kinematics the beam spin asymmetry is essentially
determined by one CFF only. Although the approximations are rather crude,
the result can serve to obtain a qualitative understanding of this GPD. The
LO analysis of the pioneering measurements of DVCS
\cite{Adletal01,Airetal01,Steetal01} on the proton suggest for $-\Delta^2
\sim 0.3\ \GeV^2$ the dominance of valence-quark GPDs with no essential
enhancement by the skewedness effect. Assuming the same features also for
the deuteron GPDs, we estimated the size of the beam spin asymmetry.


%

\end{document}